%% file: rambo-on-diet.tex
\documentclass[epj]{svjour}
\usepackage{amsmath,amssymb,amsfonts,graphicx,cite,algorithmic,algorithm,multirow}
\graphicspath{{graphics/}}
\makeatletter
\ifx\input@path\@undefined
\def\input@path{{graphics/}}
\else
\g@addto@macro\input@path{{graphics/}}
\fi
\makeatother
\newcommand{\program}[1]{\textsf{#1}}

\preprint{DESY 13-145\\ MCnet-13-10}

\title{RAMBO on diet}

\author{Simon Pl\"atzer}

\institute{DESY, Notkestrasse 85, D-22607 Hamburg, Germany}

\date{\today}

\abstract{
We describe a phase space generator which is flat for
  massless particles, and approximately flat for massive particles of
  masses much smaller than the typical mometum scales involved in the
  process. The same goal is achieved by the RAMBO algorithm, contrary
  to which our approach does only require the minimal number of
  degrees of freedom and is invertible in the sense that it provides a
  unique mapping of random numbers to the physical phase space point
  and vice versa. The very motivation to seek such an algorithm is
  explained in detail.
}

\begin{document}

\onecolumn

\maketitle

\section{Introduction}

Within applying the Monte Carlo (MC) method to cross section
computations, the phase space generation is an integral part to the
efficiency of the implementation. By phase space generation we here
refer to a particular mapping of random numbers to a physical phase
space point, excluding any methods of adaptive MC integration, but
typically implementing importance sampling in a way that the resulting
Jacobian resembles the structure of the differential cross section as
close as possible.

The very opposite to any effort of importance sampling for physical
cross sections, namely a flat population of phase space, has so far
only be addressed by Kleiss and collaborators,
\cite{Kleiss:1985gy}. Owing to the complicated structure of the
Lorentz invariant phase space manifold, this task is actually far from
trivial, though the resulting RAMBO algorithm exhibits a very
transparent and simple structure. It is therefore most often used to
provide means of cross checks for more involved phase space generator
implementations and is certainly far from being used in realistic
setups other than for this particular purpose. Besides the flat phase
space population, which will exhibit the full peak structure of the
differential cross section to the integration methods, another show
stopper in this context is certainly its exhaustive use of random
numbers, requiring $4 n$ instead of the $3 n -4$ required degrees of
freedom to generate a system of $n$ outgoing momenta.

One may therefore question why one would seek another variant of such
a flat phase space generator. We will motivate, why precisely this
property amongst others has been addressed in the present note. In
particular, we will derive a phase space generation algorithm which will
\begin{itemize}
\item perform flat\footnote{For massive particles, almost flat,
  meaning flat in the limit that all invariants are much larger than
  the masses present.} phase space population, while it will
\item require the minimum number of degrees of freedom needed for $n$
  outgoing particles, $3n-4$, and is
\item invertible in the sense that it provides a unique and invertible
  mapping between a physical phase space point and the random numbers
  used to generate that particular point.
\end{itemize}

\section{Motivation}

The motivation to seek such an algorithm originates from the adaptive
sampling of matrix element corrections for parton showers
\cite{Seymour:1994df,Norrbin:2000uu} using algorithms such as the one
implemented in the \program{ExSample} library \cite{Platzer:2011dr},
outlined in more detail in \cite{Platzer:2011bc}.
The problem we tackle here is the following: Our aim is to sample a
matrix element correction splitting kernel, which is of the form
$|{\cal M}_R(p_\perp,z,\phi;\phi_n)|^2/|{\cal M}_B(\phi_n)|^2$ with a
Born and real emission matrix elements squared ${\cal M}_{B,R}$ and is
a function of the Born phase space point $\phi_n$ and the degrees of
freedom needed for the additional emission, {\it e.g.}
$p_\perp,z,\phi$. The latter are the variables which will actually be
drawn from the associated Sudakov distribution by means of standard
methods, \cite{Platzer:2011dq}, while the Born phase space point is
fixed by the hard process considered and constitutes an external set
of parameters. Any adaptive algorithm will have to keep track of those
parameters and should do so in a way as to require the minimal number
of degrees of freedom needed to describe the Born phase space
point. Otherwise one will make the possible adaption steps more
complicated and slower convergent than needed or possible. Of course,
another choice would be to directly communicate the random numbers
from the hard process generation, which would provide the most simple
and straightforward solution to the problem at hand. Within recent
approaches to improving parton showers by fixed order matrix element
corrections \cite{Platzer:2012bs,Platzer:SubleadingMatching}, this may
not be possible anymore. Finally, we note that in the bulk of the
cases ({\it i.e.} for soft and collinear emissions), no peak structure
is present in the matrix element correction as a function of the Born
degrees of freedom. This completes our motivation and we will specify
the resulting algorithm in the next section. A generic C++
implementation of it is available publicly.

\section{The algorithm}

We consider the $n$-body phase space measure for
a total momentum $Q$ \footnote{Note that we have suppressed the
  overall factor $1/(2\pi)^{3n-4}$ for the sake of readability.},
\begin{equation}
{\rm d}\phi_n\left(\{p_1,m_1\},...,\{p_n,m_n\}|Q\right) = 
\delta\left(\sum_{i=1}^n p_n - Q\right) \prod_{i=1}^n {\rm d}^4p_i \delta(p_i^2-m_i^2)\theta(p_i^0-m_i) \ .
\end{equation}
We factor the measure by the well-known approach of iterative $1\to 2$ decays,
\begin{multline}
{\rm d}\phi_n\left(\{p_1,m_1\},...,\{p_n,m_n\}|Q\right) = 
{\rm d}\phi_n\left(\{p_1,m_1\},...,\{p_{n-2},m_{n-2}\},\{Q_{n-1},M_{n-1}\}|Q\right)\ \times\\
{\rm d}M_{n-1}^2 \times{\rm d}\phi_2\left(\{p_{n-1},m_{n-1}\},\{p_n,m_n\}|Q_{n-1}\right)  \ ,
\end{multline}
\begin{multline}
{\rm d}\phi_n\left(\{p_1,m_1\},...,\{p_n,m_n\}|Q\right) = \\
\left(\prod_{i=2}^{n} {\rm d}\phi_2\left(\{p_{i-1},m_{i-1}\},\{Q_i,M_i\}|Q_{i-1}\right)\right) \times
\left( \prod_{i=2}^{n-1}\theta\left(M_{i-1}-m_{i-1}-M_i\right)\theta\left(M_i-\sum_{k=i}^{n} m_k\right) {\rm d}M_i^2 \right)
\end{multline}
where we have made explicit the phase space limits present in the
$1\to 2$ decays, and we have identified $Q_1=Q$ as well as
$\{Q_n,M_n\}=\{p_n,m_n\}$. We express the two-body phase space
measures in the respective rest frame of the parents $Q_{i-1}$,
\begin{equation}
{\rm d}\phi_2\left(\{p_{i-1},m_{i-1}\},\{Q_i,M_i\}|Q_{i-1}\right) =
\rho(M_{i-1},M_i,m_{i-1}){\rm d}\cos\theta_{i-1}{\rm d}\phi_{i-1}
\end{equation}
where 
\begin{equation}
\rho(M_{i-1},M_i,m_{i-1}) = \frac{1}{8 M_{i-1}^2} \sqrt{(M_{i-1}^2-(M_i+m_{i-1})^2)(M_{i-1}^2-(M_i-m_{i-1})^2)}
\end{equation}
\begin{equation}
{\mathbf p}_{i-1}= - {\mathbf Q}_{i} =
4 M_{i-1} \rho(M_{i-1},M_i,m_{i-1})\left(\cos\phi_{i-1}\sin\theta_{i-1},\sin\phi_{i-1}\sin\theta_{i-1},\cos\theta_{i-1}\right)^T
\end{equation}
and the zero components are determined by the respective mass shell
conditions. We can thus perform flat phase space population, if we
generate the intermediate masses from the measure
\begin{multline}
{\rm d}M_n(M_2,...,M_{n-1}|M_1;m_1,...,m_n)= \\
\left(\prod_{i=2}^{n-1} \rho(M_{i-1},M_i,m_{i-1})
\theta\left(M_{i-1}-m_{i-1}-M_i\right)\theta\left(M_i-\sum_{k=i}^{n} m_k\right) {\rm d}M_i^2
\right)
\rho(M_{n-1},m_n,m_{n-1})
\end{multline}
in such a way as to arrive at constant weight.

\subsection{The massless case}

We shall now limit ourself to the discussion of external particles
with zero masses; non-zero masses do not pose a problem, and the
(invertible) procedure of how to obtain a set of massive momenta from
a set of massless momenta of identical total momentum is detailed in
\cite{Kleiss:1985gy}, or an alternative procedure outlined in the next section
may be used.  If all external masses vanish, $m_i=0$, the measure for
the intermediate masses is given by
\begin{equation}
{\rm d}M_n(M_2,...,M_{n-1}|M_1;0,...,0)=
\frac{1}{8^{n-1}}\prod_{i=2}^{n-1} \frac{M_{i-1}^2-M_i^2}{M_{i-1}^2} \theta(M_{i-1}^2-M_i^2)\theta(M_i^2) {\rm d}M_i^2 \ .
\end{equation}
Expressing $M_{i}= u_2\cdots u_i\ M_1$ we have
\begin{equation}
{\rm d}M_n(M_2,...,M_{n-1}|M_1;0,...,0)=
\frac{1}{8^{n-1}}M_1^{2n-4}\ 
\prod_{i=2}^{n-1} u_i^{n-1-i} (1-u_i) \theta(1-u_i)\theta(u_i){\rm d}u_i \ .
\end{equation}
Finally, substituting $v_i = (n+1-i)\ u_i^{n-i} - (n-i)\ u_i^{n+1-i}$ we arrive at the desired flat result,
\begin{equation}
{\rm d}M_n(M_2,...,M_{n-1}|M_1;0,...,0)=
\frac{1}{8^{n-1}}M_1^{2n-4} \frac{1}{(n-1)!\ (n-2)!}\ 
\prod_{i=2}^{n-1} \theta(1-v_i)\theta(v_i){\rm d}v_i \ .
\end{equation}
Performing the mass and angular integrations, we find the total phase space volume to be given by
\begin{equation}
V_n = \int {\rm d}\phi_n\left(\{p_1,0\},...,\{p_n,0\}|Q\right) = 
\left(\frac{\pi}{2}\right)^{n-1} \frac{(Q^2)^{n-2}}{(n-1)!\ (n-2)!}
\end{equation}
recovering precisely the RAMBO result.  The complete algorithm for
generating flat phase space with total momentum $Q$ from $3n-4$ random
numbers $r_1,...,r_{3n-4}$ is specified in
algorithm~\ref{algorithms:generator}, and its inverse, {\it i.e.}
solving for the random numbers $r_i$ as a function of a given set of
momenta, is given in algorithm~\ref{algorithms:hasher}.

\begin{algorithm}
\begin{algorithmic}
\STATE $Q_1\gets Q$, $M_1\gets \sqrt{Q^2}$, $M_n\gets 0$
\FOR{$i=2,...,n-1$}
\STATE solve $r_{i-1} = (n+1-i)\ u_i^{n-i} - (n-i)\ u_i^{n+1-i}$ for $u_i$
\STATE $M_i \gets u_2\cdots u_i \sqrt{Q^2}$
\STATE $\cos\theta_i\gets 2\ r_{n-5+2i}-1$, $\phi_i=2\pi\ r_{n-4+2i}$
\STATE $q_i\gets 4 M_{i-1} \rho(M_{i-1},M_i,0)$
\STATE ${\mathbf p}_{i-1}\gets q_i (\cos\phi_i\sqrt{1-\cos^2\theta_i},\sin\phi_i\sqrt{1-\cos^2\theta_i},\cos\theta_i)$
\STATE $p_{i-1}\gets (q_i,{\mathbf p}_{i-1})$, $Q_i\gets (\sqrt{q_i^2+M_i^2},-{\mathbf p}_{i-1})$
\STATE boost $p_{i-1}$ and $Q_i$ by ${\mathbf Q}_{i-1}/Q_{i-1}^0$
\ENDFOR 
\STATE $p_n\gets Q_n$
\RETURN $\{p_1,...,p_n\}$ with weight $V_n$
\end{algorithmic}
\caption{\label{algorithms:generator} The massless phase space generation algorithm.}
\end{algorithm}

\begin{algorithm}
\begin{algorithmic}
\STATE $M_1 \gets \sqrt{(p_1+...+p_n)^2}$ 
\STATE $Q_n\gets p_n$
\FOR{$i=n,...,2$}
\STATE $M_i\gets \sqrt{(p_{i}+...+p_n)^2}$, $u_i \gets M_i/M_{i-1}$
\STATE $r_{i-1}\gets (n+1-i)\ u_i^{n-i} - (n-i)\ u_i^{n+1-i}$
\STATE $Q_{i-1}\gets Q_i + p_{i-1}$
\STATE boost $p_{i-1}$ by $-{\mathbf Q}_{i-1}/Q_{i-1}^0$
\STATE $r_{n-5+2i}\gets ({\mathbf p}_{i-1}^z/|{\mathbf p}_{i-1}|+1)/2$, $\phi\gets\tan^{-1}({\mathbf p}_{i-1}^y/{\mathbf
    p}_{i-1}^x)$, $r_{n-4+2i}=(2\pi\theta(-\phi)+\phi)/(2\pi)$
\ENDFOR
\end{algorithmic}
\caption{\label{algorithms:hasher} The inverse of the massless phase space generation algorithm.}
\end{algorithm}

\subsection{The massive case}

The massive case can in principle be dealt with by the reshuffling
procedure given in the original RAMBO paper \cite{Kleiss:1985gy},
including its inverse. We here outline an alternative method, which
does not involve solving the reshuffling condition,
$\sum_i\sqrt{\xi{\mathbf p}_i^2 + m_i^2} = \sum_i |{\mathbf p}_i|$,
numerically. We start by rewriting the measure for the intermediate
masses as
\begin{equation}
{\rm d}M_n(M_2,...,M_{n-1}|M_1;m_1,...,m_n)= 
\frac{1}{8} \prod_{i=2}^{n} \frac{\rho(M_{i-1},M_i,m_{i-1})}{\rho(K_{i-1},K_i,0)}\times
\prod_{i=2}^{n-1}\frac{M_i}{K_i}\times
{\rm d}M_n(K_2,...,K_{n-1}|K_1;0,...,0)
\end{equation}
where $K_i=M_i-\sum_{k=i}^n m_k$ for $i=1,...,n-1$ and $K_n=0$. Up to
an additional weight factor the massive case is thus simply related to
the massless one upon taking care of the conversion between $K_i$ and
$M_i$. The additional weight factor will be close to unity, if all
invariants are forced into regions where they are much larger than the
masses of the external legs -- the same observation holds for the
massive variant of RAMBO.

\subsection{Numerical stability}

Having at hand the possibility to invert the phase space generation
back to the original random numbers, it is of course desirable to test
the numerical stability of the implementation. To this extent, we test
two possible sequences of both directions of the algorithm which need
to preserve their respective input values: a phase space point $\phi$
is used to calculate the random numbers associated to it which in turn
serve as input for the phase space generation step; and a given set of
random numbers is used to generate a phase space point which in turn
is inverted back to the respective random numbers. We have tested
those sequences for two and up to six massless as well as massive
outgoing particles. The typical accuracy, shown in
figure~\ref{figures:accuracy}, with per mille level number of events
with less than ten significant digits, is certainly not matching
expectations from double precision, though we argue that it is
absolutely sufficient for the purposes we have in mind. The most
probable source of the broad tails to less accurate points are the
implementation of Lorentz boosts we have used and improving their
numerical stability will be subject to future development.

\begin{figure}
\begin{center}
\input{phirphi-plot}
\input{rphir-plot}
\end{center}
\caption{\label{figures:accuracy}Left: Accuracy of a sequence of phase
  space hashing and generation from the hash, $\phi(q)\to \vec{r}\to
  \phi(p)$ for two and up to six external, massless legs. The initial
  phase space momenta $q$ have been generated with RAMBO. Right:
  Accuracy of a sequence of random number generation, phase space
  generation and phase space hashing, $\vec{r}\to \phi \to \vec{s}$
  for the same settings.  The solid lines represent a setting for a
  hadronic collider with varying center of mass energy,
  $\sqrt{\hat{s}}$, which has been sampled flat. The dashed lines
  correspond to a fixed energy collision.}
\end{figure}

\section{Summary}

We have presented a phase space generation algorithm which preforms
(almost) flat phase space generation while requiring the minimal
number of degrees present in a physical phase space point and being
invertible in the sense that it represents a unique and invertible
mapping between random numbers and momentum components. The main
application of this algorithm is in providing a measure of hashing a
phase space point to provide minimal information for adaptive
algorithms dealing with functions which are parametric in an
externally fixed set of momenta. We anticipate further application
within the context of improving the convergence of next-to-leading
order calculations carried out within the subtraction formalism.

\section*{Acknowledgments}
This work was supported by the Helmholtz Alliance ``Physics at the
Terascale''.

\appendix

\section{Availability}
\label{sections:install}

An implementation of the algorithm outlined in this note is available
as a phase space generator class within the \program{Matchbox} module
of Herwig++. A stand-alone C++ implementation is vailable upon request
from the author.

%% start contents %%%%%%%%%%%%%%%%%%%%%%%%%%%%%%%%%%%%%%%%%%%%%%%%%%%%%%%%%%%%%%
%%%%%%%%%%%%%%%%%%%%%%%%%%%%%%%%%%%%%%%%%%%%%%%%%%%%%%%%%%%%%%%%%%%%%%%%%%%%%%%%

\bibliography{rambo-on-diet}

%% end contents %%%%%%%%%%%%%%%%%%%%%%%%%%%%%%%%%%%%%%%%%%%%%%%%%%%%%%%%%%%%%%%%
%%%%%%%%%%%%%%%%%%%%%%%%%%%%%%%%%%%%%%%%%%%%%%%%%%%%%%%%%%%%%%%%%%%%%%%%%%%%%%%%
\end{document}

%% file: phirphi-plot.tex
% GNUPLOT: LaTeX picture with Postscript
\begingroup
  \makeatletter
  \providecommand\color[2][]{%
    \GenericError{(gnuplot) \space\space\space\@spaces}{%
      Package color not loaded in conjunction with
      terminal option `colourtext'%
    }{See the gnuplot documentation for explanation.%
    }{Either use 'blacktext' in gnuplot or load the package
      color.sty in LaTeX.}%
    \renewcommand\color[2][]{}%
  }%
  \providecommand\includegraphics[2][]{%
    \GenericError{(gnuplot) \space\space\space\@spaces}{%
      Package graphicx or graphics not loaded%
    }{See the gnuplot documentation for explanation.%
    }{The gnuplot epslatex terminal needs graphicx.sty or graphics.sty.}%
    \renewcommand\includegraphics[2][]{}%
  }%
  \providecommand\rotatebox[2]{#2}%
  \@ifundefined{ifGPcolor}{%
    \newif\ifGPcolor
    \GPcolortrue
  }{}%
  \@ifundefined{ifGPblacktext}{%
    \newif\ifGPblacktext
    \GPblacktexttrue
  }{}%
  % define a \g@addto@macro without @ in the name:
  \let\gplgaddtomacro\g@addto@macro
  % define empty templates for all commands taking text:
  \gdef\gplbacktext{}%
  \gdef\gplfronttext{}%
  \makeatother
  \ifGPblacktext
    % no textcolor at all
    \def\colorrgb#1{}%
    \def\colorgray#1{}%
  \else
    % gray or color?
    \ifGPcolor
      \def\colorrgb#1{\color[rgb]{#1}}%
      \def\colorgray#1{\color[gray]{#1}}%
      \expandafter\def\csname LTw\endcsname{\color{white}}%
      \expandafter\def\csname LTb\endcsname{\color{black}}%
      \expandafter\def\csname LTa\endcsname{\color{black}}%
      \expandafter\def\csname LT0\endcsname{\color[rgb]{1,0,0}}%
      \expandafter\def\csname LT1\endcsname{\color[rgb]{0,1,0}}%
      \expandafter\def\csname LT2\endcsname{\color[rgb]{0,0,1}}%
      \expandafter\def\csname LT3\endcsname{\color[rgb]{1,0,1}}%
      \expandafter\def\csname LT4\endcsname{\color[rgb]{0,1,1}}%
      \expandafter\def\csname LT5\endcsname{\color[rgb]{1,1,0}}%
      \expandafter\def\csname LT6\endcsname{\color[rgb]{0,0,0}}%
      \expandafter\def\csname LT7\endcsname{\color[rgb]{1,0.3,0}}%
      \expandafter\def\csname LT8\endcsname{\color[rgb]{0.5,0.5,0.5}}%
    \else
      % gray
      \def\colorrgb#1{\color{black}}%
      \def\colorgray#1{\color[gray]{#1}}%
      \expandafter\def\csname LTw\endcsname{\color{white}}%
      \expandafter\def\csname LTb\endcsname{\color{black}}%
      \expandafter\def\csname LTa\endcsname{\color{black}}%
      \expandafter\def\csname LT0\endcsname{\color{black}}%
      \expandafter\def\csname LT1\endcsname{\color{black}}%
      \expandafter\def\csname LT2\endcsname{\color{black}}%
      \expandafter\def\csname LT3\endcsname{\color{black}}%
      \expandafter\def\csname LT4\endcsname{\color{black}}%
      \expandafter\def\csname LT5\endcsname{\color{black}}%
      \expandafter\def\csname LT6\endcsname{\color{black}}%
      \expandafter\def\csname LT7\endcsname{\color{black}}%
      \expandafter\def\csname LT8\endcsname{\color{black}}%
    \fi
  \fi
  \setlength{\unitlength}{0.0500bp}%
  \begin{picture}(5040.00,3528.00)%
    \gplgaddtomacro\gplbacktext{%
      \csname LTb\endcsname%
      \put(1647,704){\makebox(0,0)[r]{\strut{} 1}}%
      \put(1647,1147){\makebox(0,0)[r]{\strut{} 10}}%
      \put(1647,1590){\makebox(0,0)[r]{\strut{} 100}}%
      \put(1647,2033){\makebox(0,0)[r]{\strut{} 1000}}%
      \put(1647,2476){\makebox(0,0)[r]{\strut{} 10000}}%
      \put(1647,2918){\makebox(0,0)[r]{\strut{} 100000}}%
      \put(1779,484){\makebox(0,0){\strut{}-15}}%
      \put(2063,484){\makebox(0,0){\strut{}-14}}%
      \put(2348,484){\makebox(0,0){\strut{}-13}}%
      \put(2632,484){\makebox(0,0){\strut{}-12}}%
      \put(2916,484){\makebox(0,0){\strut{}-11}}%
      \put(3201,484){\makebox(0,0){\strut{}-10}}%
      \put(3485,484){\makebox(0,0){\strut{}-9}}%
      \put(3769,484){\makebox(0,0){\strut{}-8}}%
      \put(4054,484){\makebox(0,0){\strut{}-7}}%
      \put(4338,484){\makebox(0,0){\strut{}-6}}%
      \put(481,1983){\rotatebox{-270}{\makebox(0,0){\strut{}events}}}%
      \put(3058,154){\makebox(0,0){\strut{}$\max\log_{10}(1-p^i/q^i)$}}%
    }%
    \gplgaddtomacro\gplfronttext{%
      \csname LTb\endcsname%
      \put(3351,3090){\makebox(0,0)[r]{\strut{}$n=2$}}%
      \csname LTb\endcsname%
      \put(3351,2870){\makebox(0,0)[r]{\strut{}$n=3$}}%
      \csname LTb\endcsname%
      \put(3351,2650){\makebox(0,0)[r]{\strut{}$n=4$}}%
      \csname LTb\endcsname%
      \put(3351,2430){\makebox(0,0)[r]{\strut{}$n=5$}}%
      \csname LTb\endcsname%
      \put(3351,2210){\makebox(0,0)[r]{\strut{}$n=6$}}%
    }%
    \gplbacktext
    \put(0,0){\includegraphics{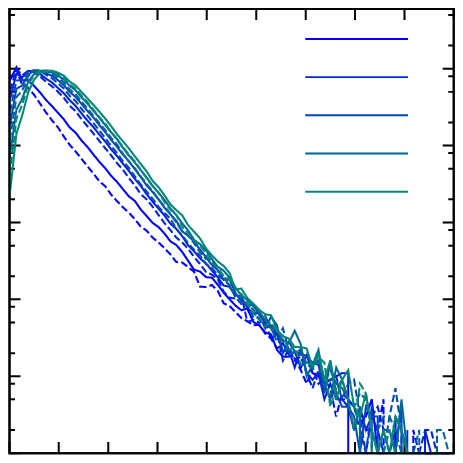}}%
    \gplfronttext
  \end{picture}%
\endgroup

%% file: rphir-plot.tex
% GNUPLOT: LaTeX picture with Postscript
\begingroup
  \makeatletter
  \providecommand\color[2][]{%
    \GenericError{(gnuplot) \space\space\space\@spaces}{%
      Package color not loaded in conjunction with
      terminal option `colourtext'%
    }{See the gnuplot documentation for explanation.%
    }{Either use 'blacktext' in gnuplot or load the package
      color.sty in LaTeX.}%
    \renewcommand\color[2][]{}%
  }%
  \providecommand\includegraphics[2][]{%
    \GenericError{(gnuplot) \space\space\space\@spaces}{%
      Package graphicx or graphics not loaded%
    }{See the gnuplot documentation for explanation.%
    }{The gnuplot epslatex terminal needs graphicx.sty or graphics.sty.}%
    \renewcommand\includegraphics[2][]{}%
  }%
  \providecommand\rotatebox[2]{#2}%
  \@ifundefined{ifGPcolor}{%
    \newif\ifGPcolor
    \GPcolortrue
  }{}%
  \@ifundefined{ifGPblacktext}{%
    \newif\ifGPblacktext
    \GPblacktexttrue
  }{}%
  % define a \g@addto@macro without @ in the name:
  \let\gplgaddtomacro\g@addto@macro
  % define empty templates for all commands taking text:
  \gdef\gplbacktext{}%
  \gdef\gplfronttext{}%
  \makeatother
  \ifGPblacktext
    % no textcolor at all
    \def\colorrgb#1{}%
    \def\colorgray#1{}%
  \else
    % gray or color?
    \ifGPcolor
      \def\colorrgb#1{\color[rgb]{#1}}%
      \def\colorgray#1{\color[gray]{#1}}%
      \expandafter\def\csname LTw\endcsname{\color{white}}%
      \expandafter\def\csname LTb\endcsname{\color{black}}%
      \expandafter\def\csname LTa\endcsname{\color{black}}%
      \expandafter\def\csname LT0\endcsname{\color[rgb]{1,0,0}}%
      \expandafter\def\csname LT1\endcsname{\color[rgb]{0,1,0}}%
      \expandafter\def\csname LT2\endcsname{\color[rgb]{0,0,1}}%
      \expandafter\def\csname LT3\endcsname{\color[rgb]{1,0,1}}%
      \expandafter\def\csname LT4\endcsname{\color[rgb]{0,1,1}}%
      \expandafter\def\csname LT5\endcsname{\color[rgb]{1,1,0}}%
      \expandafter\def\csname LT6\endcsname{\color[rgb]{0,0,0}}%
      \expandafter\def\csname LT7\endcsname{\color[rgb]{1,0.3,0}}%
      \expandafter\def\csname LT8\endcsname{\color[rgb]{0.5,0.5,0.5}}%
    \else
      % gray
      \def\colorrgb#1{\color{black}}%
      \def\colorgray#1{\color[gray]{#1}}%
      \expandafter\def\csname LTw\endcsname{\color{white}}%
      \expandafter\def\csname LTb\endcsname{\color{black}}%
      \expandafter\def\csname LTa\endcsname{\color{black}}%
      \expandafter\def\csname LT0\endcsname{\color{black}}%
      \expandafter\def\csname LT1\endcsname{\color{black}}%
      \expandafter\def\csname LT2\endcsname{\color{black}}%
      \expandafter\def\csname LT3\endcsname{\color{black}}%
      \expandafter\def\csname LT4\endcsname{\color{black}}%
      \expandafter\def\csname LT5\endcsname{\color{black}}%
      \expandafter\def\csname LT6\endcsname{\color{black}}%
      \expandafter\def\csname LT7\endcsname{\color{black}}%
      \expandafter\def\csname LT8\endcsname{\color{black}}%
    \fi
  \fi
  \setlength{\unitlength}{0.0500bp}%
  \begin{picture}(5040.00,3528.00)%
    \gplgaddtomacro\gplbacktext{%
      \csname LTb\endcsname%
      \put(1647,704){\makebox(0,0)[r]{\strut{} 1}}%
      \put(1647,1147){\makebox(0,0)[r]{\strut{} 10}}%
      \put(1647,1590){\makebox(0,0)[r]{\strut{} 100}}%
      \put(1647,2033){\makebox(0,0)[r]{\strut{} 1000}}%
      \put(1647,2476){\makebox(0,0)[r]{\strut{} 10000}}%
      \put(1647,2918){\makebox(0,0)[r]{\strut{} 100000}}%
      \put(1779,484){\makebox(0,0){\strut{}-15}}%
      \put(2063,484){\makebox(0,0){\strut{}-14}}%
      \put(2348,484){\makebox(0,0){\strut{}-13}}%
      \put(2632,484){\makebox(0,0){\strut{}-12}}%
      \put(2916,484){\makebox(0,0){\strut{}-11}}%
      \put(3201,484){\makebox(0,0){\strut{}-10}}%
      \put(3485,484){\makebox(0,0){\strut{}-9}}%
      \put(3769,484){\makebox(0,0){\strut{}-8}}%
      \put(4054,484){\makebox(0,0){\strut{}-7}}%
      \put(4338,484){\makebox(0,0){\strut{}-6}}%
      \put(481,1983){\rotatebox{-270}{\makebox(0,0){\strut{}events}}}%
      \put(3058,154){\makebox(0,0){\strut{}$\max\log_{10}(1-r^i/s^i)$}}%
    }%
    \gplgaddtomacro\gplfronttext{%
      \csname LTb\endcsname%
      \put(3351,3090){\makebox(0,0)[r]{\strut{}$n=2$}}%
      \csname LTb\endcsname%
      \put(3351,2870){\makebox(0,0)[r]{\strut{}$n=3$}}%
      \csname LTb\endcsname%
      \put(3351,2650){\makebox(0,0)[r]{\strut{}$n=4$}}%
      \csname LTb\endcsname%
      \put(3351,2430){\makebox(0,0)[r]{\strut{}$n=5$}}%
      \csname LTb\endcsname%
      \put(3351,2210){\makebox(0,0)[r]{\strut{}$n=6$}}%
    }%
    \gplbacktext
    \put(0,0){\includegraphics{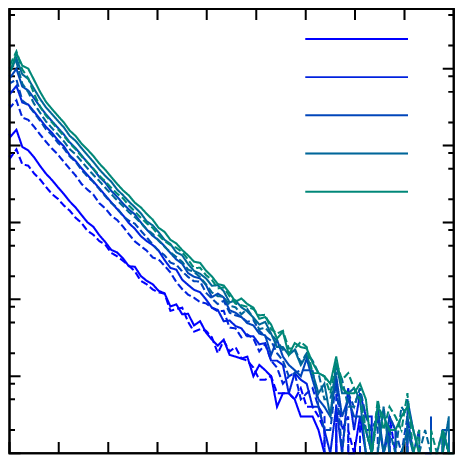}}%
    \gplfronttext
  \end{picture}%
\endgroup

%% file: rambo-on-diet.bbl
\begin{thebibliography}{1}

\bibitem{Kleiss:1985gy}
R.~Kleiss, W.~J. Stirling, and S.~D. Ellis,
\newblock Comput. Phys. Commun. {\bf 40}, 359 (1986).

\bibitem{Seymour:1994df}
M.~H. Seymour,
\newblock Comp. Phys. Commun. {\bf 90}, 95 (1995), hep-ph/9410414.

\bibitem{Norrbin:2000uu}
E.~Norrbin and T.~Sj{\"o}strand,
\newblock Nucl. Phys. {\bf B603}, 297 (2001), hep-ph/0010012.

\bibitem{Platzer:2011dr}
S.~Platzer,
\newblock Eur.Phys.J. {\bf C72}, 1929 (2012), 1108.6182.

\bibitem{Platzer:2011bc}
S.~Platzer and S.~Gieseke,
\newblock Eur.Phys.J. {\bf C72}, 2187 (2012), 1109.6256.

\bibitem{Platzer:2011dq}
S.~Platzer and M.~Sjodahl,
\newblock Eur.Phys.J.Plus {\bf 127}, 26 (2012), 1108.6180.

\bibitem{Platzer:2012bs}
S.~Platzer,
\newblock (2012), 1211.5467.

\bibitem{Platzer:SubleadingMatching}
S.~Platzer,
\newblock in preparation  (2013).

\end{thebibliography}
